\documentclass[10pt]{article}
\usepackage[margin=1in]{geometry}
\usepackage[numbers]{natbib}
\usepackage{pifont}
\usepackage{geometry}
\usepackage{fleqn}
\usepackage{graphicx}
\usepackage{booktabs}
\usepackage{txfonts}
\usepackage{amssymb}
\usepackage{subfig}
\usepackage{color}
\usepackage{ulem}
\usepackage{float}
\usepackage{hyperref}
\usepackage{miller}

\title{\textbf{A Robust Machine Learned Interatomic Potential for Nb: Collision Cascade Simulations with accurate Defect Configurations}}
\author{%
  Utkarsh Bhardwaj\textsuperscript{1}, 
  Vinayak Mishra\textsuperscript{1,2}, 
  Suman Mondal\textsuperscript{1}, 
  Manoj Warrier\textsuperscript{1,2} \\[1ex]
  \textsuperscript{1}\textit{Computational Analysis Division, BARC, Visakhapatnam, Andhra Pradesh, India – 530012} \\[0.5ex]
  \textsuperscript{2}\textit{Homi Bhabha National Institute, Anushaktinagar, Mumbai, Maharashtra, India – 400094} \\[0.5ex]
}

\date{}

\begin{document}

\maketitle

\begin{abstract}
Niobium (Nb) and its alloys are extensively used in various technological applications owing to their favorable mechanical, thermal and irradiation properties. Accurately modeling Nb under irradiation is essential for predicting microstructural changes, defect evolution, and overall material performance.  Traditional interatomic potentials for Nb fail to predict the correct self-interstitial atom (SIA) configuration, a critical factor in radiation damage simulations. We develop a machine learning interatomic potential (MLIP) using the Spectral Neighbor Analysis Potential (SNAP) framework, trained on \textit{ab-initio} Density Functional Theory (DFT) calculations, which accurately captures the relative stability of different SIA dumbbell configurations. The resulting potential reproduces DFT-level accuracy while maintaining computational efficiency for large-scale Molecular Dynamics (MD) simulations. Through a series of validation tests involving elastic, thermal, and defect properties---including collision cascade simulations---we show that our SNAP potential resolves persistent limitations in existing Embedded Atom Method (EAM) and Finnis--Sinclair (FS) potentials and is effective for MD simulations of collision cascades. Notably, it accurately captures the ground-state SIA configuration of Nb in the primary damage of a collision cascade, offering a robust tool for predictive irradiation studies.

\end{abstract}

\section{Introduction}

Niobium (Nb) and its alloys play critical roles in high-temperature and radiation-intensive environments, such as nuclear reactors and aerospace industries, due to their high melting point and desirable mechanical properties~\cite{wojcik1993processing, english1984physical}. Molecular Dynamics (MD) is an effective simulation method for predicting radiation damage in materials~\cite{Stoller2012293, Zinkle201265}. However, the accuracy of an MD simulation is limited by the interatomic potential used to calculate energies and forces. Traditional interatomic potentials (e.g., EAM, FS) for Nb have struggled to reproduce correct defect properties, especially the ground-state configuration of self-interstitial atoms (SIAs)~\cite{lin2017molecular, lee2001second, zhang2016experimental, fellinger2010force}.

In body-centered cubic (BCC) non-magnetic metals, the most common ground state SIA configuration is a $\langle 111\rangle$ dumbbell, while $\langle 110\rangle$ is less stable with higher formation energy~\cite{ma2019universality}. Some group-6 metals, such as Cr, Mo, and W, may exhibit symmetry-broken SIA configurations, but for Nb, $\langle 111\rangle$ remains consistently the lowest-energy orientation~\cite{ma2019symmetry}. The discrepancy between different interatomic potentials regarding the morphology of defects is a well studied topic in material damage studies \cite{Bhardwaj3IAPCmp, warrier5IAPCmp}. Furthermore, SIA morphology strongly influences elasticity, stress fields, and defect mobility, ultimately altering microstructural evolution under high-dose irradiation~\cite{OSETSKY200065, OSETSKY2002852, ma2019universality}. Consequently, accurately predicting the stable SIA orientation is crucial for reliable large-scale simulations of Nb under defect-inducing environments.

Machine Learning Interatomic Potentials (MLIPs) offer a promising alternative to classical potentials by approximating the potential energy surface with high fidelity, while remaining less computationally expensive than first-principles methods~\cite{Mortazavi2023}. MLIPs have shown success in capturing mechanical, thermal, and defect properties of transition metals more accurately than classical approaches~\cite{okita2022construction, byggmastar2020gaussian}. The robustness of the MLIP needs a validation on a long simulation that involves all the complexities of the desired application. Nevertheless, despite close matches with DFT data, MLIPs can exhibit instabilities during extended MD simulations~\cite{vita2023data}. It is thus vital to validate MLIPs under application-specific scenarios, particularly those involving complex phenomena such as high-energy irradiation.


In this work, we develop and extensively validate a SNAP-based MLIP~\cite{thompson2015spectral} for Nb using a robust DFT dataset generated via VASP~\cite{kresse1996efficiency}. The subset used to train the potential is curated using diversity scoring and optimization methods \cite{friedman2023the, doptimal}. Our primary goal is to accurately capture Nb’s mechanical, thermal, and defect characteristics, particularly under irradiation-induced damage. By benchmarking against well-known EAM~\cite{Fellinger2010} and FS~\cite{Mendelev2022} potentials, we demonstrate that the developed SNAP potential resolves the persistent discrepancy in existing potentials (e.g., incorrect $\langle 110 \rangle$ SIA ground state~\cite{lin2017molecular}) by reproducing the DFT-observed $\langle 111 \rangle$ SIA dumbbell orientation, but also remains stable and predictive during 5\,keV collision cascade simulations. These simulations reveal that differences in defect formation energies lead to significantly distinct primary damage morphologies, underscoring the MLIP’s suitability for large-scale irradiation studies of Nb.

\section{Methodology}

\subsection{DFT Dataset Generation}
We generated an extensive DFT dataset to ensure that the relevant atomic configurations were adequately sampled:
\begin{itemize}
    \item \textbf{Strained Configurations:} Covering a range of elastic deformations around the equilibrium lattice parameter.
    \item \textbf{Defect Configurations:} Including single vacancies up to the size four and self-interstitial atoms (SIAs) in different configurations.
    \item \textbf{Liquid Configurations:} Sampling disordered states, which are relevant for exploring formation of complex defect morphologies in a collision cascade ~\cite{bhardwaj2024identifying}.
\end{itemize}

All DFT calculations were performed using VASP~\cite{kresse1996efficiency}, employing projector augmented-wave (PAW) pseudopotentials~\cite{kresse1999ultrasoft} and a Perdew--Burke--Ernzerhof (PBE) exchange-correlation functional. The computed energies, forces, and stresses formed the training targets for the potential. We preform ab-initio MD with the defect and liquid configurations and extract multiple frames from each of the simulations. 

\begin{figure}[H]
\centering
\includegraphics[width=.6\textwidth]{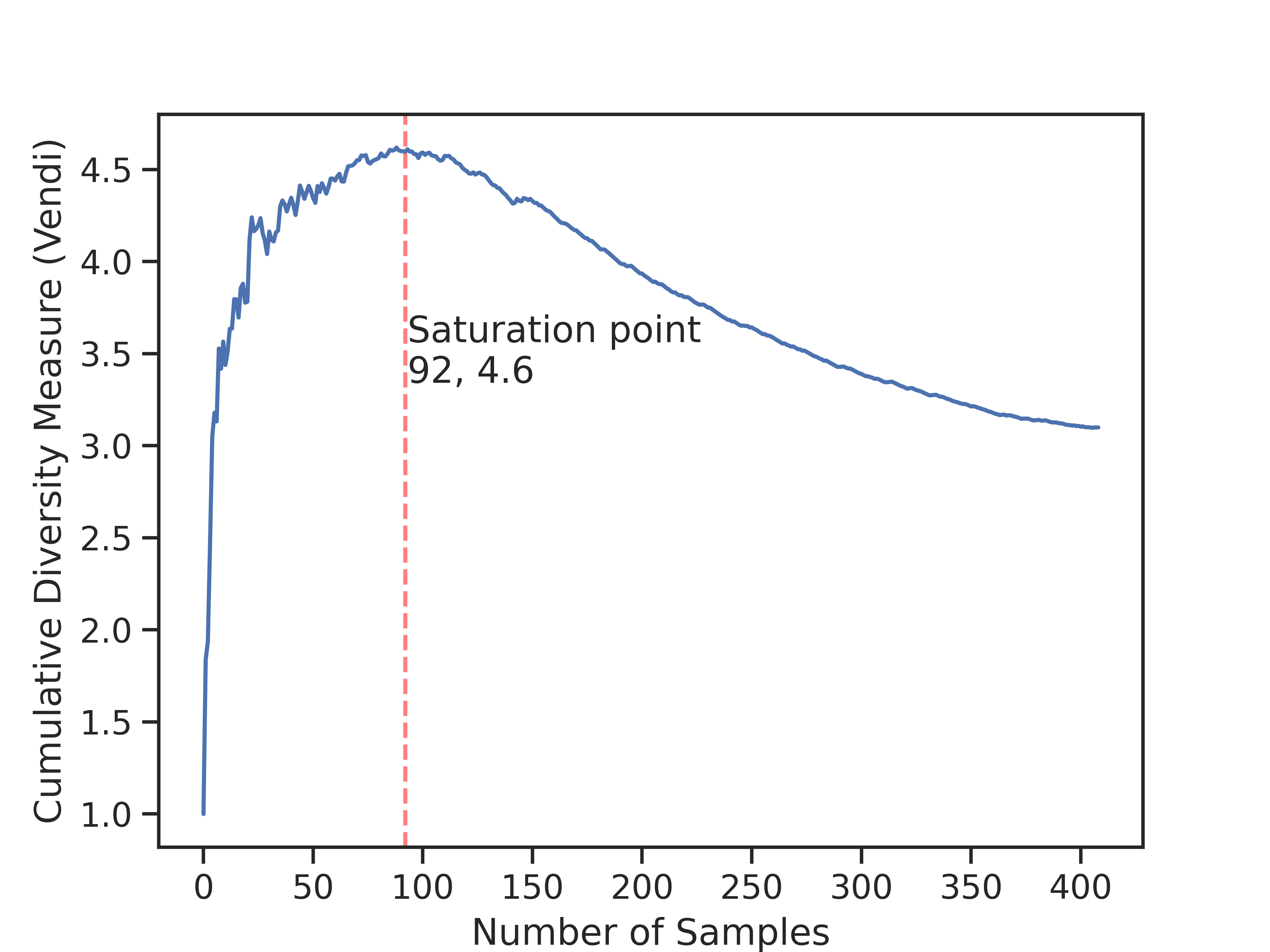}
\caption{\label{Fig-vendi} The Cumulative diversity scores based on Vendi method \cite{friedman2023the} for DFT dataset of category SIA defects. We chose the saturation point of the score to decide the maximum number of datasets to be chosen for a category.
}
\end{figure}

To enhance the effectiveness of our training dataset, we employed the Vendi Score \cite{friedman2023the}, a diversity evaluation metric defined as the exponential of the Shannon entropy of the eigenvalues of similarity matrix of the features of the dataset. The features for dataset selection algorithm included energy per atom, maximum force and mean force in addition to the Bispectral features representing atomic configuration. This ensures that the diversity is measured in terms of energies, forces as well as the configuration space. Fig~\ref{Fig-vendi} shows the cumulative diversity score as we keep on adding the samples to the dataset. The absolute value of the diversity score shows the effective number of samples in the dataset while we chose the maximum saturation point as the measure for maximum number of samples that we would use from a category of configurations. This approach allowed us to assess the diversity within each category of configurations and determine the appropriate number of samples accordingly. These scores were then combined with a D-optimal selection method~\cite{doptimal, deldossi2022sub} to select the most representative configurations. The D-optimal selection method seeks to maximize the determinant of the information matrix, thereby minimizing the variance of parameter estimates.  We train the potential on 550 total configurations out of which 90\% are selected in the training set and 10\% are kept for testing and validation. 

\subsection{Machine Learning Potential Fitting}
We used the SNAP framework~\cite{Zuo2019, rohskopf2023fitsnap}, which leverages bispectral feature descriptors derived from atomic neighbor environments via spherical harmonics. These feature descriptors capture the necessary radial and angular correlations of the atomic configuration. Because the bispectral features have linear relationship with energy, they can be efficiently trained with simple linear models. The continuous closed form of bispectral features ensures that the forces can be trained using the derivatives of the feature descriptor. Bispectral features are rotationally invariant and provide a rich representation of the local chemical environment. SNAP also seamlessly integrates the ZBL~\cite{Ziegler85} pair potential for short-range repulsion, ensuring faithful modeling of high-energy pair-wise collisions. The SNAP framework's accuracy, transferability and computational efficiency has enabled its widespread adoption for developing reliable MLIPs across diverse materials~\cite{bideault2024polyvalent}.

The SNAP framework smoothly combines the short-range repulsion using ZBL~\cite{Ziegler85} pair potential for correctly modelling the close atomic interactions essential for simulating collision cascades for irradiation. Because of these qualities of the Bispectral features and SNAP framework, they are used extensively to train accurate and robust MLIPs for various materials and alloys \cite{bideault2024polyvalent} 

\begin{figure}[H]
\centering
\includegraphics[width=\textwidth]{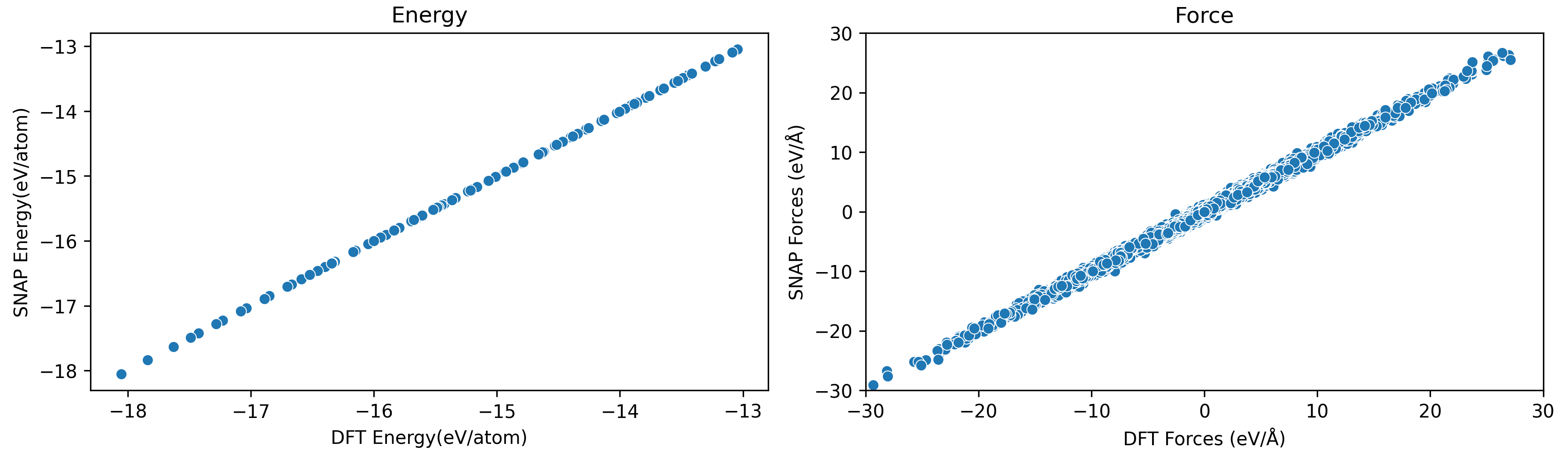}
\caption{\label{Fig0} The SNAP potential predictions for (a) Energy and (b) Forces against the reference DFT values.
}
\end{figure}

Fig.\ref{Fig0} shows the energy and force predictions of the developed MLIP against the reference DFT values. A Bayesian optimization scheme \cite{optuna_2019} was employed to tune key hyper-parameters such as the band limit (chosen to be \texttt{jmax=6}) and radial cut-off (chosen to be \texttt{rcut=5.11}), balancing model accuracy with computational efficiency. The mean absolute errors (MAEs) on validation/test set for our final model are 1.21$\times 10^{-5}$\,eV/atom (energy), 6.19$\times 10^{-5}$\,eV/\AA{} (force), and 7.46$\times 10^{-8}$\,eV/\AA$^3$ (stress). 

\subsection{Validation and Comparison}
Following training, we validated the SNAP potential against a variety of properties:
\begin{itemize}
    \item \textbf{Elastic Properties:} Comparison with DFT, experiments, and two reference potentials (EAM~\cite{Fellinger2010} and FS~\cite{Mendelev2022}).
    \item \textbf{Thermal Properties:} Bulk modulus at finite temperature, thermal expansion coefficient, and approximate melting point.
    \item \textbf{Defect Properties:} Vacancy and SIA formation energies, emphasizing the orientation of the stable dumbbell configuration.
\end{itemize}

Additionally, we performed 5\,keV collision cascade simulations using LAMMPS~\cite{Thompson2022} in ten random directions for statistics. The EAM potential was stiffened with a ZBL potential~\cite{Ziegler85} in the short-range region, while the FS potential was already pre-stiffened. An $80\times80\times80$ cubic simulation box with periodic boundaries was equilibrated at 300\,K before initiating cascades via a primary knock-on atom (PKA). The outermost unit cells of the simulation box are fixed while the adjacent two unit cells from the boundary are temperature controlled at 300 K. We then analyzed defects and dumbbell orientations using the Csaransh software suite~\cite{BhardwajJOSS} that employs Savi algorithm~\cite{BhardwajSavi} for detailed defect morphology analysis.

\section{Results}

\begin{figure}[H]
\centering
\includegraphics[width=\textwidth]{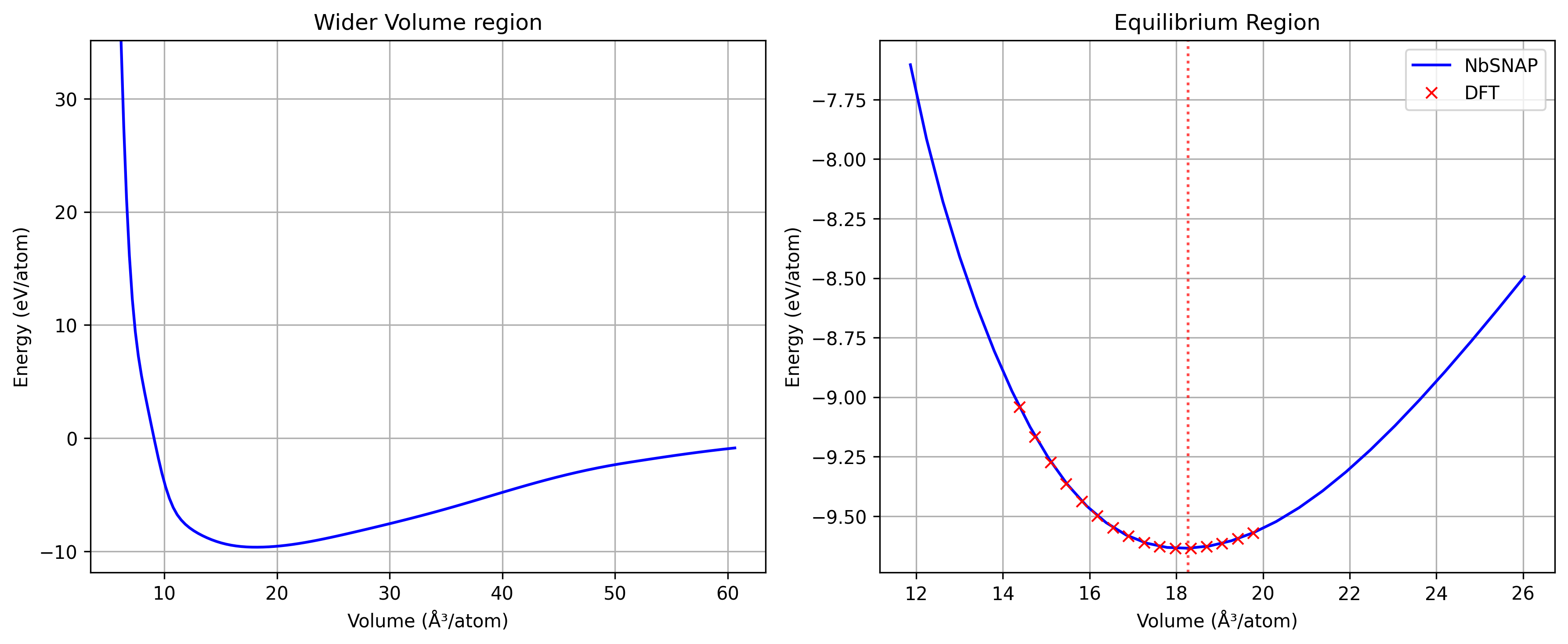}
\caption{\label{Fig1} (a) Energy--volume curve for Nb over a broad volume range, and (b) a focused view around the equilibrium region, comparing the SNAP potential with DFT.}
\end{figure}

Figure~\ref{Fig1} demonstrates a good agreement in energy-volume curves around equilibrium. We find a smooth transition from the repulsive regime, through the equilibrium, and then to the attractive part with a gradual decay.

\subsection{Elastic and Thermal Properties}
Table~\ref{tab:Elastic} summarizes the elastic constants, lattice parameter, melting point, and thermal expansion coefficient for the developed SNAP potential, benchmarked against DFT, experiment, and two reference potentials. Our SNAP potential closely matches DFT for $C_{11}, C_{12},$ and the bulk modulus. Although $C_{44}$ from DFT is lower than the experimental value---a known trend in some DFT calculations~\cite{nguyen2006self, ma2019universality, ma2019symmetry}---the SNAP potential reproduces this lower $C_{44}$ accurately. The thermal expansion coefficient is calculated by running a NPT simulation at different temperatures for 0.5 ns and observing the change in the volume of the simulation box. The melting point is calculated via liquid-solid coexistence simulations.

\begin{table}[!htbp]
\centering
\caption{Elastic and Thermal Properties of Nb with Relative Errors (\%) with respect to DFT or Experimental Values~\cite{rumble2019crc} whichever is less.}
\label{tab:Elastic}
\begin{tabular}{@{}lccccc@{}}
\toprule
\textbf{Property} & \textbf{SNAP} & \textbf{FS} & \textbf{EAM} & \textbf{DFT} & \textbf{Exp.} \\
\midrule
Lattice Param. (Å)      & 3.31 (0.0\%)   & 3.31 (0.0\%)   & 3.31 (0.0\%)   & 3.31   & 3.30      \\
$C_{11}$ (GPa)          & 236  (-0.4\%)  & 247  (0.4\%)   & 233  (-1.7\%)  & 237    & 246       \\
$C_{12}$ (GPa)          & 135  (0.7\%)   & 132  (-1.5\%)  & 123  (-8.2\%)  & 134    & 135       \\
$C_{44}$ (GPa)          & 14   (0.0\%)   & 47   (67\%)     & 32   (14.3\%)  & 14    & 28        \\
Bulk Modulus (GPa)      & 169  (0.0\%)   & 170.7 (0.2\%)  & 160.3 (-5.1\%) & 169    & 170.3     \\
Shear Modulus (GPa)     & 29   (0.0\%)   & 47.4 (58\%)     & 32.1 (10.7\%)  & 29     & 30        \\
Poisson's Ratio         & 0.40 (2.6\%)   & 0.35 (-7.89)     & 0.34 (-10.5)  & 0.39   & 0.38--0.40 \\
Melting Point (K)       & 2800 (---)     & 3050 (---)     & 3000 (---)     & --     & 2750       \\
Thermal Exp. ($10^{-6}$K$^{-1}$) & 9.8 (22\%) & 9.6 (20\%) & 9.6 (20\%) & 8.0 & 7.3 \\
\bottomrule
\end{tabular}
\end{table}

\subsection{Defect Formation Energies}
\begin{figure}[H]
\centering{\includegraphics[width=\textwidth]{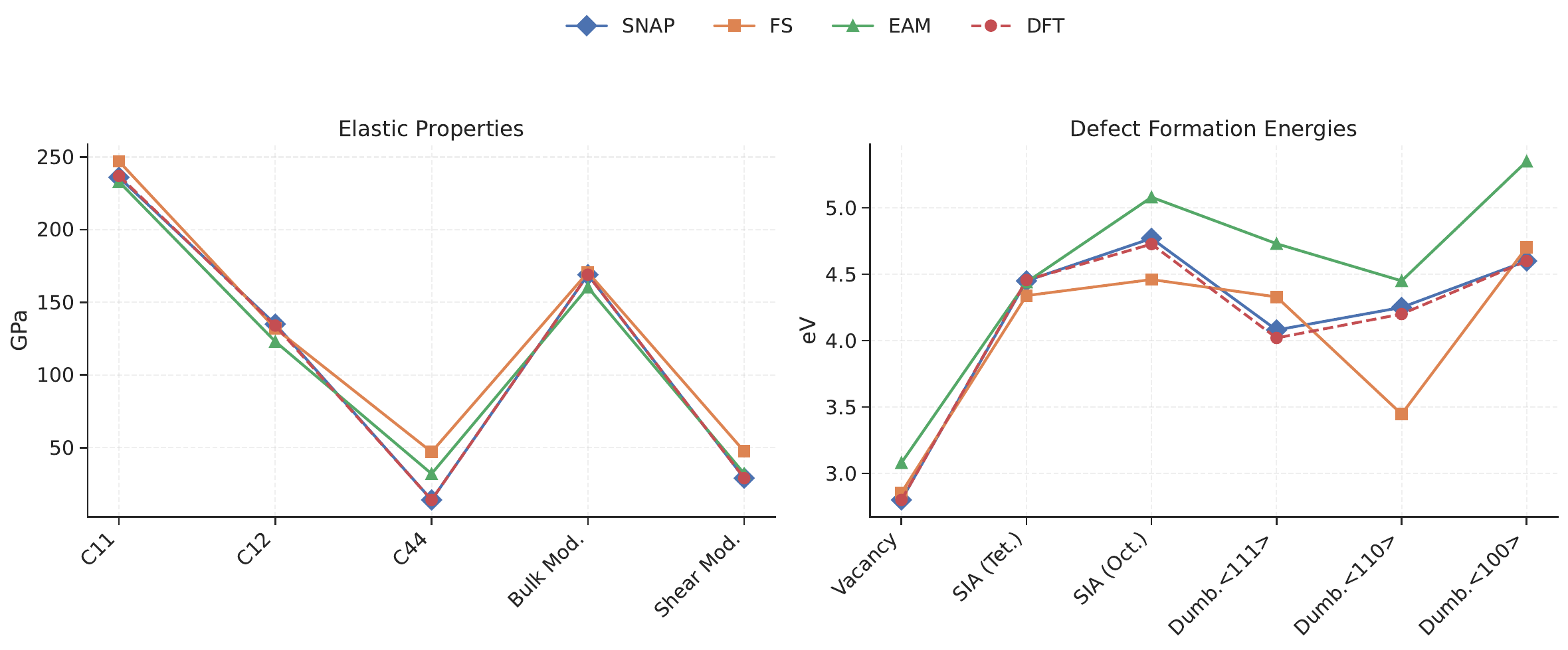}}
\caption{\label{Fig2} Comparison of (a) elastic properties and (b) defect formation energies calculated from different potentials and DFT. Both EAM and FS favor $\langle 110\rangle$ as the lowest-energy SIA, while DFT and SNAP correctly predict $\langle 111\rangle$ for Nb.
}
\end{figure}

Table~\ref{tab:Defect} lists formation energies for a single vacancy and various self-interstitial configurations. Notably, our SNAP potential agrees well with the DFT result by identifying $\langle 111\rangle$ as the lowest-energy dumbbell orientation. In contrast, EAM and FS incorrectly favor $\langle 110\rangle$, underscoring a fundamental discrepancy in classical potentials. Fig.\ref{Fig2} shows a comparative summary of elastic and defect properties for Nb calculated using different potentials with reference to DFT calculations. 

\begin{table}[!htbp]
\centering
\caption{Defect Formation Energies (in eV) with Relative Errors (\%) with respect to DFT.}
\label{tab:Defect}
\begin{tabular}{@{}lccccc@{}}
\toprule
\textbf{Defect Property} & \textbf{SNAP} & \textbf{FS} & \textbf{EAM} & \textbf{DFT} & \textbf{Exp.} \\
\midrule
Vacancy Formation    & 2.80 (0.0\%)   & 2.85 (1.8\%)   & 3.08 (10.0\%)  & 2.80 & 2.60--3.00 \\
SIA: Tetrahedral     & 4.45 (-0.1\%)  & 4.34 (-2.6\%)  & 4.44 (-0.4\%)  & 4.46 & --         \\
SIA: Octahedral      & 4.77 (0.9\%)   & 4.46 (-5.6\%)  & 5.08 (7.5\%)   & 4.73 & --         \\
SIA: Dumbbell $\langle 111\rangle$ & 4.08 (1.5\%)   & 4.33 (7.7\%)   & 4.73 (17.7\%) & 4.02 & -- \\
SIA: Dumbbell $\langle 110\rangle$ & 4.25 (1.2\%)   & 3.45 (-17.9\%) & 4.45 (6.0\%)  & 4.20 & -- \\
SIA: Dumbbell $\langle 100\rangle$ & 4.60 (0.0\%)   & 4.70 (2.2\%)   & 5.35 (16.3\%) & 4.60 & -- \\
\bottomrule
\end{tabular}
\end{table}

\subsection{Irradiation Studies}
\begin{figure}[H]
\centering
\includegraphics[width=\textwidth]{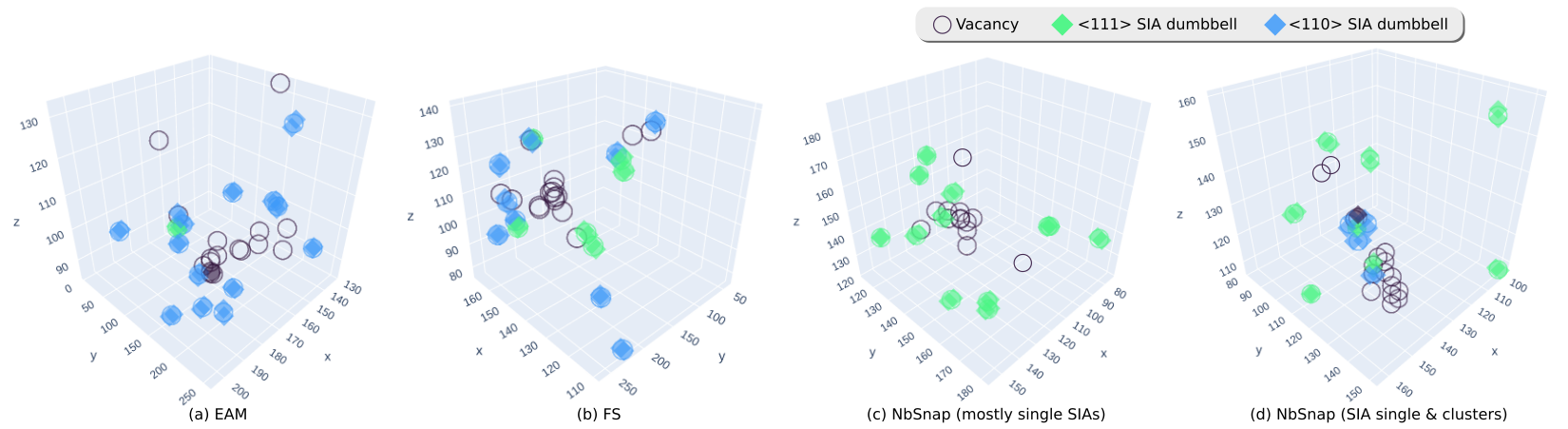}
\caption{\label{Fig3} Primary damage states in 5\,keV collision cascades simulated using different potentials. Single SIA defects in EAM and FS primarily adopt $\langle 110\rangle$ orientation, whereas SNAP yields $\langle 111\rangle$. Both orientations appear in clustered SIAs.}
\end{figure}

To evaluate the performance of our SNAP potential under irradiation, we carried out 5\,keV collision cascade simulations at 300\,K. Figure~\ref{Fig3} illustrates the final damage state for the SNAP, FS, and EAM potentials. Although each potential produces a similar overall number of residual defects, the single SIA dumbbell orientations differ markedly: SNAP predominantly yields $\langle 111\rangle$ dumbbells, whereas FS and EAM favor $\langle 110\rangle$.

Table~\ref{tab:Cascade1} compares the overall defect statistics, showing that the mean number of total defects and clusters is broadly comparable. The SNAP MLIP displays a slightly higher tendency for SIA to cluster.

\begin{table}[!htbp]
\centering
\caption{Primary Damage Properties from 5\,keV Collision Cascade Simulations (averaged over ten runs).}
\label{tab:Cascade1}
\begin{tabular}{@{}lccc@{}}
\toprule
\textbf{Property}        & \textbf{SNAP} & \textbf{FS} & \textbf{EAM} \\
\midrule
Mean \# of Defects       & 10.1 & 11.4 & 10.8 \\
Mean \# of Clusters      & 3.4  & 2.5  & 2.5  \\
Max. Vacancy Cluster Size & 10   & 10   & 9    \\
Max. SIA Cluster Size    & 5    & 3    & 3    \\
\bottomrule
\end{tabular}
\end{table}

\begin{figure}[H]
\centering
\includegraphics[width=.6\textwidth]{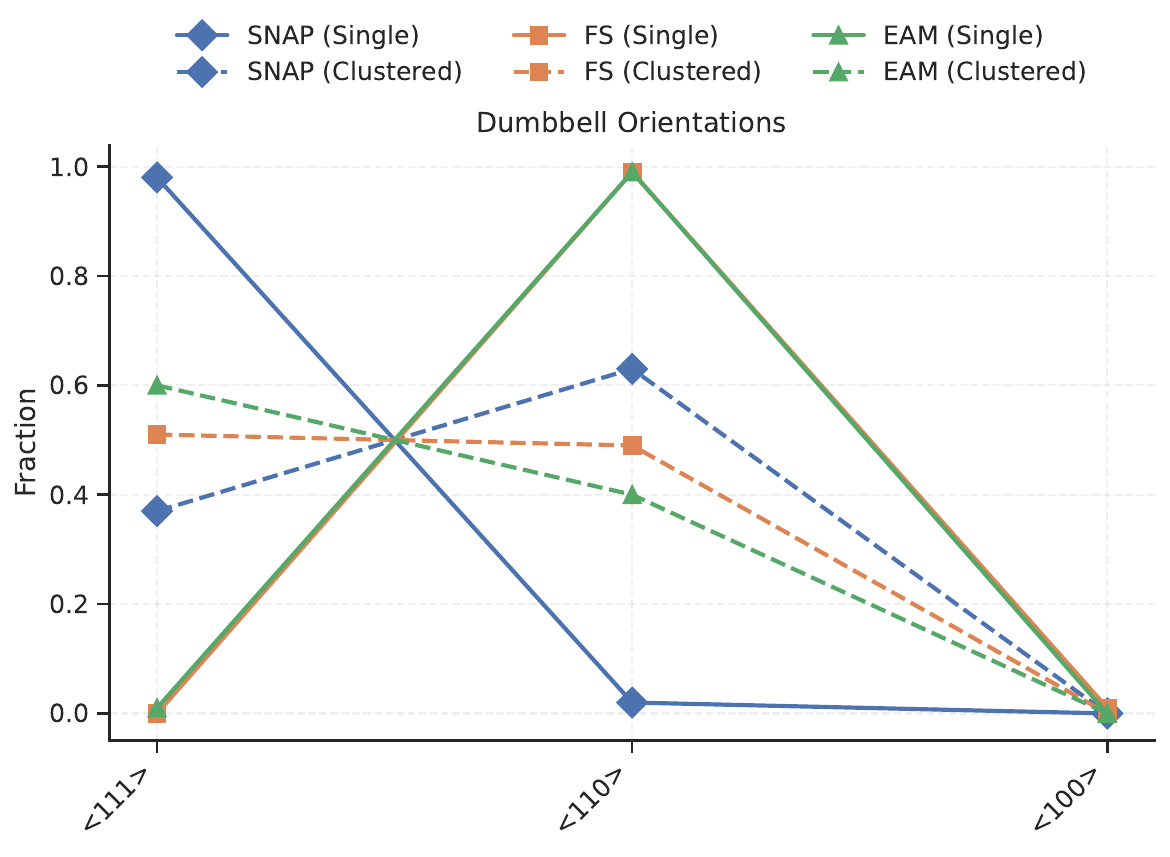}
\caption{\label{Fig4} Fraction of single and clustered dumbbells present in the primary damage produced using the three potentials. The values are averaged for the ten different 5keV collision cascades. The snap potential produces $\langle 111 \rangle$ single SIA dumbbells while the other two potentials produce $\langle 110 \rangle$.}

\end{figure}

Table~\ref{tab:Cascade2} reports the orientations of SIA dumbbells in the primary damage state. Consistent with DFT ground state configuration, the SNAP MLIP predominantly yields $\langle 111\rangle$ single dumbbells in the primary damage, whereas FS and EAM yield $\langle 110\rangle$. This highlights the impact of ground-state SIA configurations on the primary damage predictions. Fig~\ref{Fig4} highlights the differences in the defect orientations between the potentials. We observed that the clustered SIAs adopt both $\langle 111 \rangle$ and $\langle 110  \rangle$ orientations. The small clusters in BCC materials are known to form various different morphologies including di-interstitial C-15 like rings and their basis where the dumbbell orientations tend to be $\langle 110 \rangle$ \cite{BhardwajSavi, BhardwajClassify}. The developed potential includes liquid configurations in the training dataset. The presence of liquid dataset in the training process of a potential enables the potential to produce energetically favourable complex defect morphology during a collision cascade \cite{bhardwaj2024dpa, warrier5IAPCmp, Bhardwaj3IAPCmp}. The developed potential also produces a few ring like clusters with $\langle 110 \rangle$ SIA dumbbells as shown in Fig~\ref{Fig3}(d).


\begin{table}[!htbp]
\centering
\caption{Dumbbell Orientations in the Primary Damage State of Collision Cascades (Averaged Percentages).}
\label{tab:Cascade2}
\begin{tabular}{@{}lccc@{}}
\toprule
\textbf{Property}                & \textbf{SNAP} & \textbf{FS}  & \textbf{EAM} \\
\midrule
\% $\langle 111\rangle$ Single Dumbbells       & 98   & 0    & 1    \\
\% $\langle 110\rangle$ Single Dumbbells       & 2    & 99   & 99   \\
\% $\langle 100\rangle$ Single Dumbbells       & 0    & 1    & 0    \\
\% $\langle 111\rangle$ Clustered Dumbbells    & 37   & 51   & 60   \\
\% $\langle 110\rangle$ Clustered Dumbbells    & 63   & 49   & 40   \\
\% $\langle 100\rangle$ Clustered Dumbbells    & 0    & 0    & 0    \\
\bottomrule
\end{tabular}
\end{table}

\section{Discussion}

The results emphasize how the SIA formation energies influence the defect configurations in the primary damage in a collision cascade. The differences in defect configurations in primary damage state affects diffusion, clustering, and defect interactions, ultimately shaping microstructural evolution under high-dose radiation~\cite{OSETSKY200065, OSETSKY2002852}. Compared to EAM and FS, our SNAP potential matches DFT more closely for mechanical and other properties. Although MLIPs can exhibit instabilities over long simulations~\cite{vita2023data}, the ten successful 5\,keV collision cascade runs validate the robustness of our SNAP model. With the inclusion of liquid configurations in our potential training, the collision cascades also produce some complex ring like SIA cluster morphologies (Fig~\ref{Fig3}) that have been earlier shown to be stable in Fe and W \cite{LiuC15, DEZERALD2014219, BhardwajClassify, bhardwaj2024dpa, warrier5IAPCmp}. By capturing the correct defect configurations in the primary damage, our MLIP provides a reliable basis for studying extended phenomena such as void swelling, radiation-induced creep and embrittlement.

\section{Conclusion}

We present a SNAP-based ML potential for Nb that accurately reproduces key mechanical, thermal, and defect properties, including the correct SIA ground state. We used advanced data selection algorithms to optimize the dataset used for training the potential. Through testing on 5\,keV collision cascades, we confirm its stability and fidelity under conditions relevant to irradiation, matching expected behaviors from DFT and experimental insights. This resolves persistent discrepancies found in traditional EAM and FS potentials and ensures strong transferability for large-scale simulations to model radiation damage, defect and microstructural evolution in Nb.

\section*{Acknowledgements}
We acknowledge the HPC team at CAD for the HPC maintenance and for support in installing LAMMPS.

\bibliographystyle{unsrt}
\bibliography{content}

\vspace{0.5cm}
\section*{Credit Authorship Statement}
\begin{itemize}
    \item \textbf{Utkarsh Bhardwaj}: Machine Learning potential fitting, Validation, Writing
    \item \textbf{Vinayak Mishra}: DFT data curation and generation, DFT data processing
    \item \textbf{Suman Mondal}: Molecular Dynamics simulations, Literature survey on Nb potentials
    \item \textbf{Manoj Warrier}: Conceptualization, Technical Discussions, Supervision
\end{itemize}

\end{document}